Supercontinuum Generation in the Low-Peak-Power Regime Using a Single-Stage Multipass Cell

# Supercontinuum Generation in the Low-Peak-Power Regime Using a Single-Stage Multipass Cell

Zeyang Hu, Kilian Fritsch, and Oleg Pronin[1]
*n2-Photonics GmbH, Hans-Henny-Jahnn-Weg 53, 22085 Hamburg, Germany*[a)]

(*Electronic mail: zeyang.hu@n2-photonics.de)

Multipass-cell (MPC) based spectral broadening is a robust approach for ultrashort pulse compression. Its extension to oscillator-level pulse energies is fundamentally limited by insufficient nonlinear phase accumulation. We demonstrate a single-stage MPC system driven directly by a 400 nJ, 4 W, 118 fs, 10 MHz oscillator that delivers efficient spectral broadening from 800 to 1200 nm with >75% transmission. The generated spectrum supports a Fourier-limited pulse duration of 8.1 fs, indicating excellent compressibility at MW-level input peak power. These results establish a feasible operating regime for spectral broadening in a single-stage MPC at a low MW-level input peak power using a bulk Kerr nonlinear medium. This compact, high-repetition-rate source holds strong potential for multiphoton microscopy and coherent Raman imaging.

## I. INTRODUCTION

Pulse compression and spectral broadening have become essential methods in a wide range of scientific and technological fields[1]. Besides their role in basic research, these techniques are now widely used in applied and industrial scenarios, including microscopy[2–5] and spectroscopy[6–8], where high sensitivity and stability are often required. These techniques generate ultrashort pulses with broadened spectra, improving detection sensitivity and advancing high-resolution optical measurements.

Large mode area (LMA) photonic crystal fibers (PCFs) sources are an established method for SC generation. SC generation in LMA fibers is primarily driven by self-phase modulation (SPM) and optical wave breaking (OWB). Increasing the mode area reduces optical intensity and nonlinear effects, thereby helping suppress self-focusing and heat accumulation and enabling higher-power operation without exceeding the material damage threshold[9]. Südmeyer et al. 10 first extended spectral broadening to high average power (18 W) by employing LMA PCFs, generating 33-fs pulses. However, enlarging the mode area, while allowing higher power transmission, limits spectral broadening, highlighting the trade-off between fiber damage and nonlinear performance[11]. In 2018, LMA-12 PCFs (NKT Photonics) were tested in Ref. 12. This fiber had a mode field diameter (MFD) of 10.3 $\mu$m at 1064 nm and a length of 8-10 cm. It generated spectra from 800 nm to 1200 nm when pumped by a Yb:YAG thin-disk oscillator with a peak power of 4.6 MW. However, surface damage was observed on the LMA-12 fibers even at input powers around the damage threshold. In the same year, Shen et al. 13 investigated LMA-15 PCFs with a core diameter of 15.1 $\mu$m and an MFD of 12.8 $\mu$m at 1222 nm. A 5-m-long fiber was pumped by a picosecond Nd:$YVO_4$ laser (1064 nm, 7 ps, 1 MHz) with 3.8 W pump power corresponding to a peak power of approximately 0.54 MW. The fiber produced a broad spectrum extending from 600 nm to 1600 nm. However, the authors reported that 3.8 W pump power was already close to the damage threshold of the LMA-15 fiber, mainly due to loss and heating near the bend edge. Although these fibers are effective for spectral broadening, their micrometer-scale core size requires precise alignment to ensure efficient coupling of the pump laser into the fiber: Seidel et al. 12 reported coupling efficiencies of 65–80% for the tested fibers; thus, the LMA-12 fiber can be reasonably expected to exhibit a coupling efficiency of approximately 70–75%; LMA-15 tested by Shen et al. 13 only had coupling efficiency at around 60%. Both Ref. 12 and Ref. 13 required specific optical setups (e.g., focusing lenses) to mode match the beam in order to allow efficient coupling into the fiber. In practical situations, the coupling condition is highly sensitive to external perturbations, and even slight drifts in the pump laser beam induced by vibrations, thermal fluctuations, or mechanical instability can lead to a noticeable degradation of coupling efficiency. Such variations not only induce unwanted intensity fluctuations at the fiber input but also accumulate during propagation, leading to an increase in the relative intensity noise (RIN) of the generated supercontinuum [14–16].

The Herriott-type MPC-based spectral broadening approach stands for its simplicity and scalability[17], requiring only two curved mirrors and a Kerr medium. More importantly, an MPC can tolerate small mode mismatches[18] without inducing additional transmission losses or optical damage, thereby enhancing system robustness and reducing daily maintenance requirements. In MPC-based spectral broadening, SPM is enhanced compared to single-pass by increasing the effective nonlinear interaction length through multiple beam passes within the nonlinear medium. Nevertheless, reports of pulse compression to sub-10 fs durations are still few[19–23]. This results from the challenge of maintaining the required group delay dispersion (GDD) bandwidth[24] and controlling the dispersion introduced by the mirrors, which, in turn, reduces the effectiveness of the spectral broadening. Moreover, losses introduced by dielectric mirrors and the propagation medium (especially solid media[23]) in such designs are also non-negligible. Beyond these dispersion- and loss-related limits, the achievable spectral broadening is ultimately constrained by the peak power available within the

[a)]http://n2-photonics.de/

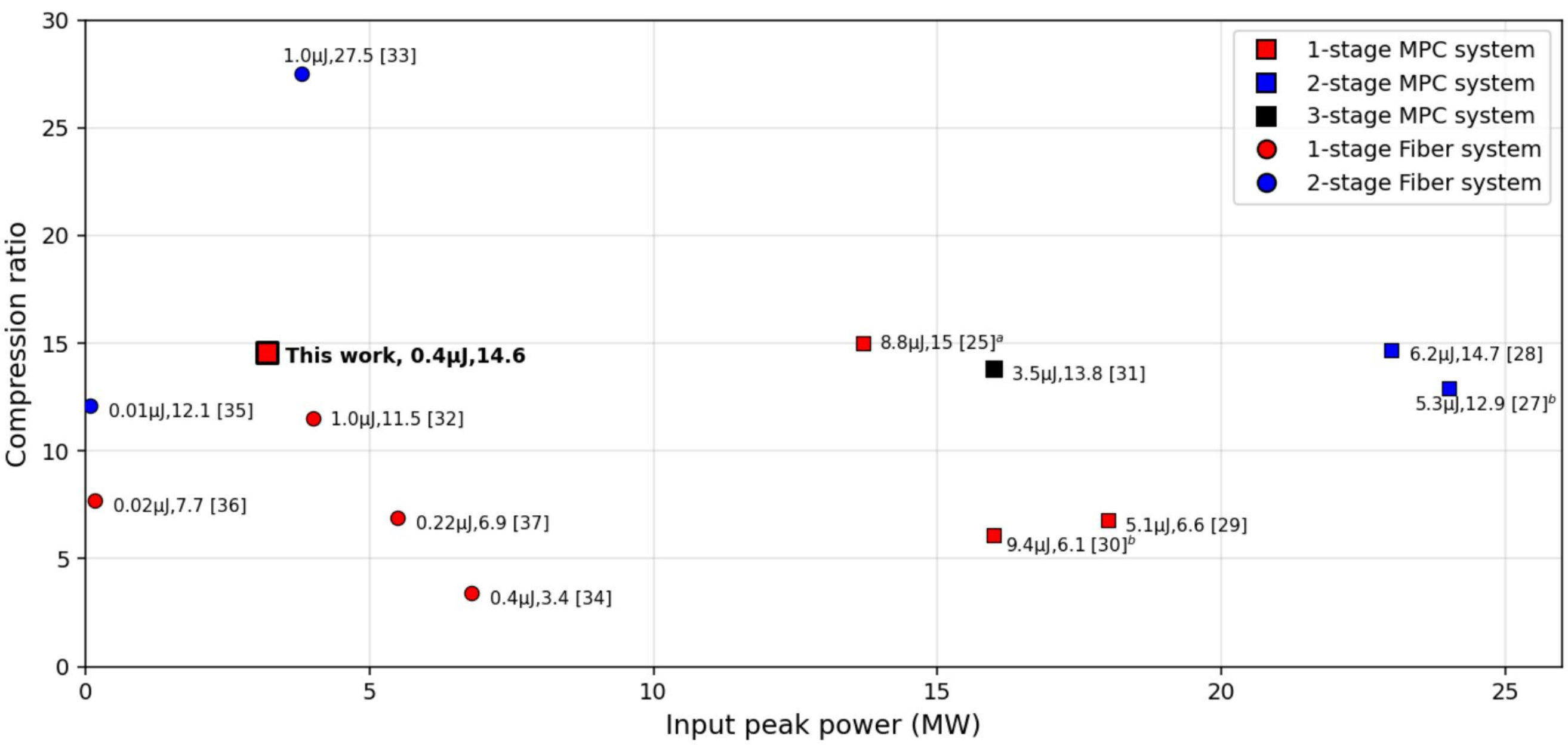

FIG. 1. Overview of compression results at 1030 nm for input energy lower than 10 μJ. The color encodes the spectral broadening mechanism (fiber or bulk based MPC), and the annotations indicate the input pulse energy and compression ratio.
[a]This work was done with a double-pass MPC compressor.
[b]Only the compression achieved by the MPC scheme is considered for consistency.

TABLE I. Overview of MPC systems with sub-10 fs pulse durations[a]

| | Units | This paper | 2nd stage of | | | | |
|---|---|---|---|---|---|---|---|
| | | | [19] | [20] | [21] | [22] | [23] |
| $P_{\mathrm{pk,input}}$ | MW | **3.2** | 280 | $30\times10^3$ | $178\times10^3$ | $25\times10^3$ | $1.6\times10^3$ |
| $P_{\mathrm{pk,output}}$ | MW | **37**[b] | $1.0\times10^3$ | $6.0\times10^4$[c] | $3.0\times10^5$ | $2.8\times10^4$ | $2.9\times10^3$ |
| $\tau_{\mathrm{input}}$ | fs | **118** | 39 | 31 | 50 | 32 | 46 |
| $\tau_{\mathrm{output}}$ | fs | **8.1**[d] | 7 | 7 | 9.6 | 13 | 8.2 |
| $f_{\mathrm{rep}}$ | MHz | **10** | 1 | 0.5 | 0.001 | 0.1 | 1 |
| Transm. | % | **75** | 91 | 82 | 74 | 46 | 75 |
| Medium | – | **FS**[e] | Ar | Ar, He | Ne | Kr | FS |
| Compression ratio | – | **14.6** | 5.6 | 4.4 | 5.2 | 2.5 | 5.6 |

[a]Peak power, $P_{\mathrm{pk}}$; pulse durations, $\tau$; average powers, $P_{\mathrm{avg}}$; overall transmission calculated as system input/system output, Transm.; repetition rate, $f_{\mathrm{rep}}$. [b]Calculated based on output energies and spectra supported Fourier-transform-limited (FTL). [c]The output peak power for Ref. [20] is an upper-bound estimate obtained by linearly scaling measurements on a sampled beam to the full pulse energy. [d]The spectrally broadened pulse supports a FTL duration of 8.1 fs. [e]FS: Fused Silica.

MPC[25]. At low input pulse energies, the reduced peak power limits nonlinear phase accumulation, making it difficult to drive sufficient self-phase modulation without substantially increasing the interaction length or the thickness of the nonlinear medium. Ref. 26 (Fig. 4) illustrates that bulk MPC systems are typically operated in the μJ regime, whereas sub-μJ compression is mainly achieved with fibers. The sub-μJ regime for bulk MPC remains largely unexplored (for typical ultrafast laser systems delivering compressed pulses in the few-hundred-femtosecond range, sub-μJ pulse energies correspond to peak powers in the MW regime).

In this work, we investigate the low MW-level peak power range for SC in MPCs. The main motivation is to apply MPC broadening to low peak power laser systems, especially oscillators. With only 400 nJ input energy (3.2 MW peak power at 10 MHz), the system generated broadband output spanning 800–1200 nm with over 75% transmission efficiency. The Fourier Transform Limited (FTL) pulse corresponding to the measured spectrum has a full width at half maximum (FWHM) duration of approximately 8.1 fs, demonstrating the capability of this compact, low-energy MPC scheme to deliver high-quality ultrashort pulses.

Figure 1 highlights two distinct compression regimes for input pulse energies below 10 μJ[25,27–37]. Our work shifts MPC-based compression into the sub-μJ domain while maintaining a relatively high compression ratio. Table 1 summarizes a comparison between our system and other reported sub-10 fs MPC-based systems[19–23]. It is worth noting that previous MPC-based systems generally relied on two-stage configurations. This two-stage design was mainly due to the long seed pulse duration, which required pre-compression to boost the peak power before final compression. As a result, the second stage operated at much higher input peak powers, enabling the strongest nonlinear interaction and final few-cycle pulse compression. To enable a fair comparison, we benchmark our results against the second stage of the reported systems, which is responsible for the main nonlinear phase accumulation. Under this comparison, our system delivers a compression ratio of 14.6 (based on FTL), which is higher than those reported systems, although the corresponding input peak powers in those studies are a few orders of magnitude larger. Our system shows that efficient few-cycle generation is possible without extreme peak powers.

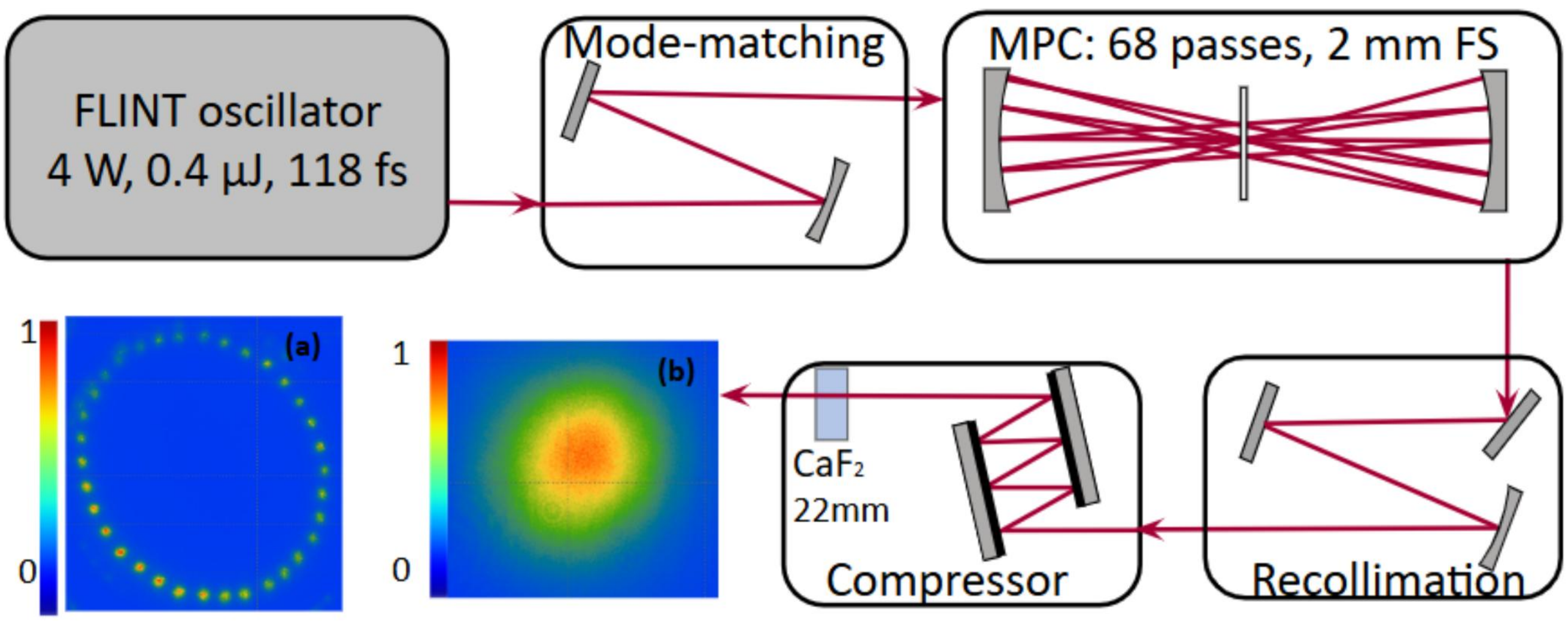

FIG. 2. Schematic of the nonlinear broadening and pulse compression (FS: Fused Silica) (a) One of the MPC patterns on a CCD camera. (b) Output beam profile after MPC stage.

## II. EXPERIMENTAL SETUP

The experimental setup is illustrated in Figure 2. A commercial oscillator (FLINT, Light Conversion) served as the driving source, delivering 118 fs pulses at a repetition rate of 10 MHz with 0.4 μJ pulse energy, corresponding to an average power of 4 W. Assuming a Gaussian temporal pulse profile, this corresponds to an input peak power of 3.2 MW. The laser output was adjusted to provide a peak power suitable for driving the setup, which was designed to achieve a spectral bandwidth of 800–1200 nm. The laser output was mode-matched to the MPC's eigenmode using a telescope. The Herriott cell consisted of two highly reflective mirrors with a radius of curvature of 50 mm (both have -50 $fs^2$ of GDD), separated by 97 mm. The chosen mirror separation and curvature place the MPC well within its stability region, providing a well-defined eigenmode for all passes. In this configuration, the beam reflected 68 passes within the MPC, and a 2 mm-thick AR-coated fused silica was placed at the center of the MPC. A thicker fused silica plate was avoided because it would introduce additional positive dispersion, leading to temporal pulse broadening and a reduction in peak power, thereby degrading the nonlinear spectral broadening. After 68 passes through the fused silica plate, the accumulated B-integral reached 27.3 rad (0.4 rad per pass). The MPC was operated in ambient air, where the total B-integral contribution from air was negligible, approximately $4 \times 10^{-3}$ rad. The system maintained 75% transmission, primarily limited by losses in the AR-coated fused silica and the dispersive mirrors[23]. The feasibility of pulse compression to 10 fs was investigated separately, as described in Section IV.

## III. SPECTRAL BROADENING RESULTS AND OUTPUT BEAM QUALITY

Figure 3(a) shows the spectra obtained in our system. The output spectrum after MPC (red) is significantly broadened, extending from 800 nm to 1200 nm. The logarithmic representation (dashed lines) further emphasizes the efficient generation of weak spectral components at the spectral edges, which are less visible on the linear scale (the spectrometer exhibits a noise floor at around $10^{-3}$ on the log scale). The temporal profile retrieved from the linear spectrum is shown in Figure 3(b). The FTL pulse has a FWHM of approximately 8.1 fs, confirming that the generated spectrum is sufficient to support sub-10 fs pulses once residual dispersion is fully compensated.

Figure 2(b) shows the measured output beam profile after the MPC stage. Elliptical fitting of the beam indicates diameters of approximately 4.2 mm (Dx) and 4.0 mm (Dy) along the axes, with an ellipticity of 0.94, confirming that the output beam remains nearly circular with no visible signs of distortion. The beam quality was further measured to be $M^2_{eff}$ = 1.107 ± 0.014 (ISO11146-2), indicating that the output beam is close to an ideal Gaussian profile (Figure 4). It should be noted that, due to the 1100 nm upper sensitivity limit of the Si sensor, the beam quality was verified only within this spectral range.

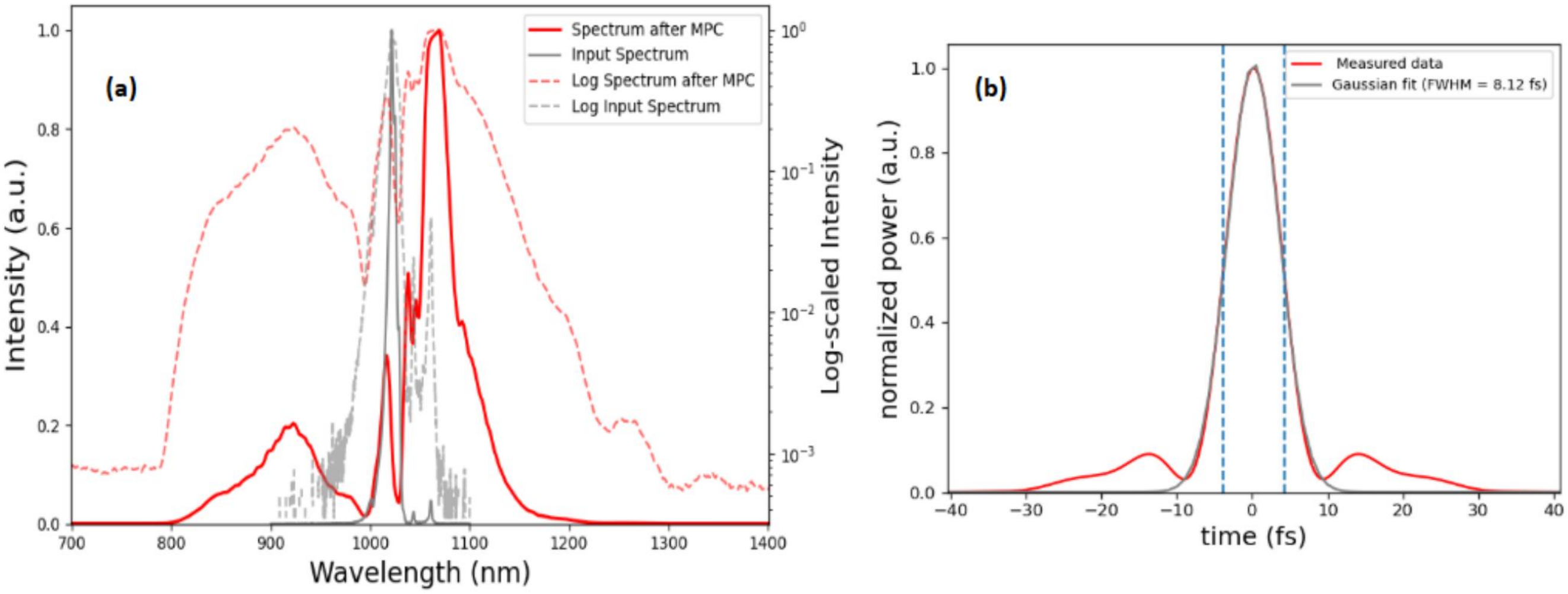

FIG. 3. (a) Spectral characteristics of the MPC output. (b) FTL temporal profile retrieved from the output spectrum (Fig. 3(a)): Pulse duration of 8.1 fs compared with a Gaussian fit.

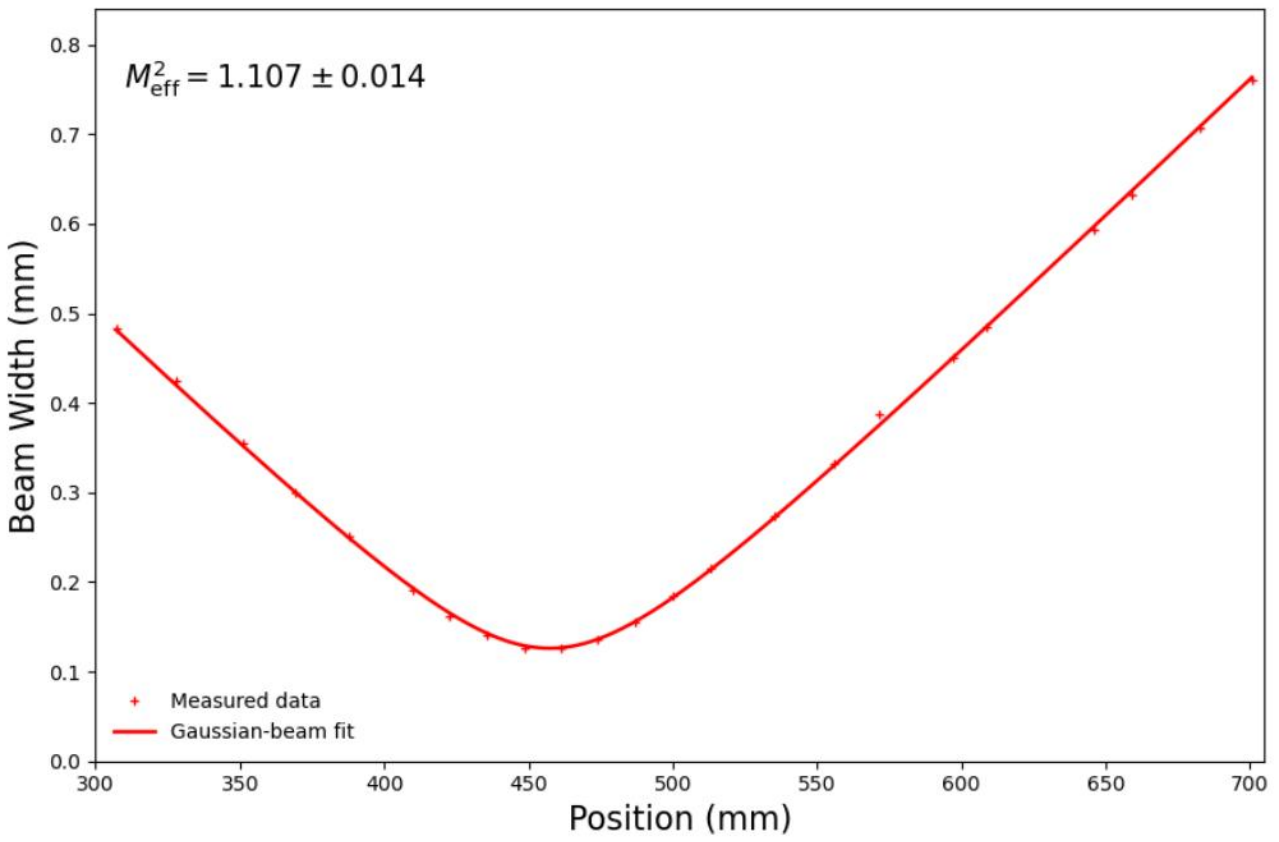

FIG. 4. Measurement of beam quality of M2. The central wavelength of 1030 nm is chosen for the measurement.

Experimentally, the output beam quality remains close to a fundamental Gaussian profile ($M_2 \approx 1$). As reported in Ref. 38, a preserved M² value already indicates reasonable homogeneity. Therefore, this suggests that the nonlinear interaction in our MPC is reasonably uniform across the beam profile. Moreover, the nonlinear phase accumulated per pass is kept moderate ( 0.4 rad, to our knowledge, bulk-MPC systems reported in the literature roughly follow the original design guideline that limits the B-integral per pass in the Herriott cell to below $\frac{\pi}{5}$[18]). A comprehensive spatio-temporal characterization will be addressed in future work.

## IV. NONLINEAR COMPRESSION RESULTS

Unfortunately, the FLINT oscillator was available only for a limited period, which prevented the completion of the compression experiments. Instead, a comparable setup was implemented using a regenerative amplifier system attenuated to deliver 0.4 W average power (0.4 µJ, 1 MHz, 90 fs) from a Pharos UP laser (Light Conversion). The main goal was to verify the feasibility of the compression. The setup used the same MPC mirrors, but separated by 98 mm, a 2 mm-thick AR-coated fused silica plate was positioned 4 mm from the MPC center. After 58 passes through the fused silica plate, the accumulated B-integral reached 20.4 rad (0.35 rad per pass). Pulse characterization was done with a D-max device (Sphere

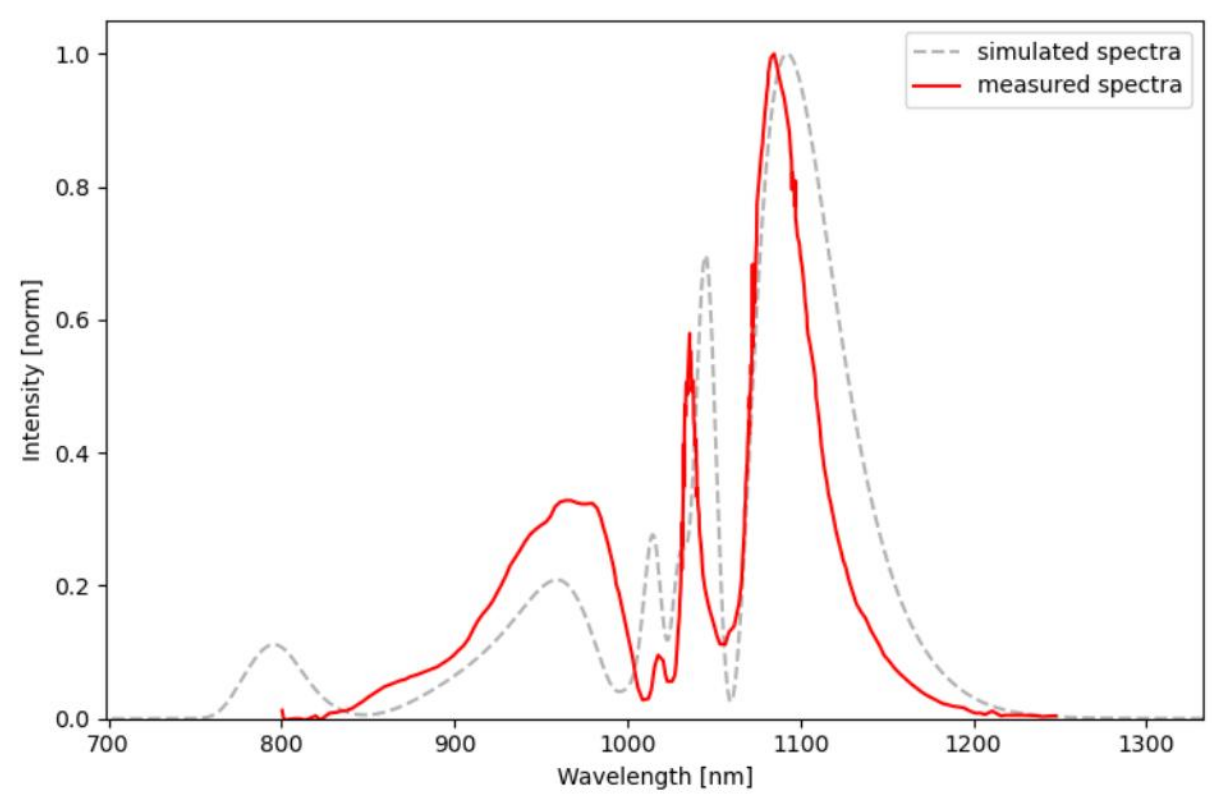

FIG. 5. Comparison between simulated and measured spectra.

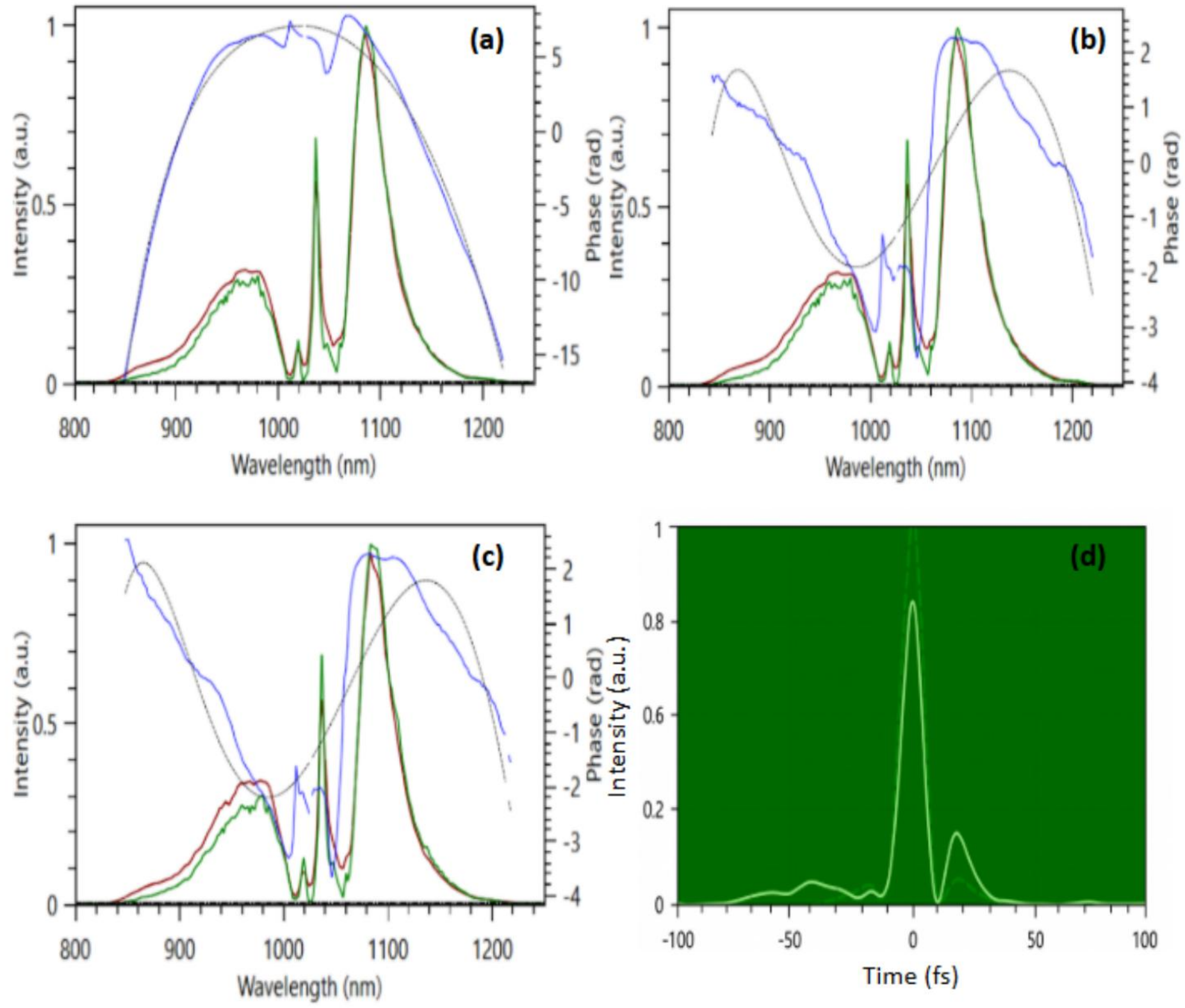

FIG. 6. (a)-(c) D-max retrieval results in the spectral domain. Red line: Measured linear spectra. Green line: Retrieved linear spectra. Blue line: Retrieved spectral phase. Black line: 4th order polynomial fit of the spectral phase. (a) After the MPC output (before the compressor). (b) After the MPC with the simulated compressor in D-max software. (c) After the actual compressor (CaF + chirped mirrors). (d) Temporal intensity compared to FTL of the spectra in Fig.6 (c). Dashed green: Temporal intensity profile of the transform-limited pulse-10.32 fs. Light Green: Temporal intensity profile of the measured pulse-10.0 fs). Relative peak power: 78.5%.

Ultrafast Photonics). The comparison between the measured and simulated spectra is presented in Figure 5. In our simulation, an input power of 0.37 W (0.37 μJ, 1 MHz, 100 fs) was used, incorporating an average mirror reflection loss of 0.3% per bounce and a medium transmission loss of 0.2% to account for experimental conditions. The slightly reduced input power accounts for propagation losses. The result shows a good agreement between the simulated and experimental spectra.

Figure 6 shows D-max retrieval results in the spectral domain. In Figure 6(a), the D-max retrieval taken after the MPC indicates a distorted phase with $-3 \times 10^2$ $fs^2$ of GDD and $+1.8 \times 10^3$ $fs^3$ of TOD, indicating the pulse accumulated strong dispersion during the MPC propagation. A numerical simulation of the compressor was available in the D-max software (Figure 6(b), indicating that a compressed pulse could be achieved using CaF and chirped mirrors. Figure 6(c) shows the experimentally retrieved result after the actual compressor, which was in good agreement with the simulated result, confirming successful dispersion compensation. The pulse duration was reduced to around 10 fs with a relative peak power compared to FTL is 78.5% (Figure 6(d), the retrieval with a root-mean-square error of 0.6%), very close to the results shown in Figure 3(b). The presence of the pedestal suggests incomplete dispersion compensation, which will be further investigated in future work.

## V. CONCLUSION

We demonstrated an efficient single-stage MPC compressor directly driven by an oscillator. Operating at only 400 nJ (3.2 MW peak power, 10 MHz repetition rate), the system achieved spectral broadening from 800 to 1200 nm with 75% transmission. The output spectrum corresponds to a FTL pulse duration of 8.1fs, indicating excellent compressibility. The FLINT oscillator was used solely to demonstrate spectral broadening, whereas the Pharos UP amplifier was employed to experimentally verify the feasibility of pulse compression. Compared with previously reported MPC systems, our setup operates at 10 times lower peak powers while still delivering few-cycle pulses with high beam quality. This highlights that efficient sub-10 fs pulse generation can be realized without extreme GW-level peak powers, thus offering a practical alternative to LMA fibers. Due to its compact footprint, and excellent output characteristics, the setup holds strong potential for demanding applications in nonlinear microscopy, as well as in other areas where ultrashort, high-repetition-rate light sources are required.

## ACKNOWLEDGMENTS

We would like to thank the Light Conversion team, namely Valdas Maslinskas and Mantvydas Mikulis, for providing the lasers that enabled these remarkable measurements. We also thank Igor Rebrov and Tomin Joy for valuable discussions, and Christian Franke for designing the optomechanical components.

## DATA AVAILABILITY STATEMENT

Data underlying the results presented in this paper are not publicly available at this time but may be obtained from the authors upon reasonable request.

## AUTHOR DECLARATIONS

### Conflict of Interest

The authors declare a conflict of interest. The work was funded by n2-Photonics GmbH, which commercializes the pulse-shortening technology. The company was co-founded by Oleg Pronin and Kilian Fritsch.

### Author Contributions

**Zeyang Hu**: Conceptualization (supporting); Data curation (lead); Formal analysis (lead); Investigation (lead); Methodology (equal); Software (equal); Validation (lead); Visualization (lead); Writing – original draft (lead); Writing – review & editing (equal). **Kilian Fritsch**: Conceptualization (supporting); Data curation (supporting); Formal analysis (supporting); Investigation (equal); Software (lead); Supervision (supporting); Validation (supporting); Visualization (supporting); Writing – review & editing (equal). **Oleg Pronin**: Conceptualization (lead); Data curation (supporting); Formal analysis (supporting); Investigation (supporting); Methodology (equal); Project administration (lead); Software (equal); Supervision (lead); Validation (supporting); Visualization (equal); Writing – review & editing (equal).